\documentclass[runningheads]{llncs}
\usepackage[T1]{fontenc}
\usepackage{cite}
\usepackage{algorithm}
\usepackage{algpseudocode}
\usepackage{chemformula}
\usepackage[version=4]{mhchem}
\usepackage{soul}
\usepackage{enumitem}
\usepackage{bm}
\usepackage{cancel}
\usepackage{graphicx}
\usepackage{subcaption}
\usepackage{multirow}
\usepackage{amsmath}
\usepackage{hyperref}
\hypersetup{
    colorlinks=true,
    linkcolor=blue,
    filecolor=magenta,      
    urlcolor=cyan,
    pdftitle={Overleaf Example},
    pdfpagemode=FullScreen,
    }

\usepackage{graphicx}
\newcommand{\ie}{\emph{\i.e.}}
\newcommand{\eg}{\emph{\eg.}}
\newcommand{\etc}{\emph{\etc.}}

\begin{document}
%
\title{Counterexample Generation for Infinite-State Chemical Reaction Networks}
%
\titlerunning{Counterexample Generation for Infinite-State Chemical Reaction Networks}
%
\author{Mohammad Ahmadi\inst{1} \and
Zhen Zhang\inst{2}\and
Chris Myers \inst{3} \and
Chris Winstead \inst{2} \and
Hao Zheng\inst{1}}

\authorrunning{M. Ahmadi et al.}
%
\institute{University of South Florida, Tampa FL, USA \and
Utah State University, Logan UT,USA \and
University of Colorado, Boulder CO, USA \\
\email{\{mahmadi,haozheng\}@usf.edu}\\
\email{\{zhen.zhang,chris.winstead\}@usu.edu}\\
\email{chris.myers@colorado.edu}
}

%

%
\maketitle              
\begin{abstract}
Counterexample generation is an indispensable part of model checking process.
In stochastic model checking, counterexample generation is a challenging problem as it is not enough to find a single trace that violates the given property. Instead, a potentially large set of traces with enough probability to violate the property needs to be found.
This paper considers counterexample generation for chemical reaction network (CRN) models with potentially infinite state space.
A method based on bounded model checking using SMT solving is developed for counterexample generation for CRNs.
It intends to find a small set of property violating paths of a given model such that they collectively have a total probability that is above a given threshold.
A unique challenge is due to the highly connected state space of CRNs where a counterexample is only a tiny subset of all property violating paths.
To address such challenges, this paper presents a number of optimizations including a divide-and-conquer technique to scale up the counterexample generation method for large CRN models. 
This paper reports results from experiments on a number of infinite-state CRN models. 
\end{abstract}
\section{Introduction}
Model checking provides the capability to automatically prove or refute that a model exhibits a set of properties. 
A striking feature of model checking is that it can generate a counterexample for efficient debugging in the case that the given property is refuted. 
These capabilities are extremely valuable in verifying critical system designs, and therefore model checking has become a prominent verification technique.

For performance and dependability analysis, the target systems often show stochastic behavior, and are usually modeled using probabilistic formalisms such as discrete-time Markov chains (DTMCs) or continuous-time Markov chains (CTMCs), \emph{etc}. 
Probabilistic model checkers such as PRISM \cite{kwiatkowska2011prism} and more recently STORM\cite{dehnert2017storm} are used to verify such models against properties specified in a probabilistic logic like PCTL\cite{hansson1994logic} or CSL\cite{aziz1996verifying}.
These model checkers calculate the probability of certain events on a model by solving a system of linear equations\cite{kwiatkowska2007stochastic}. 
The numerical nature of these model checking algorithms renders them incapable of generating a counterexample at state space level when a property is refuted.
In the case of a property refutation, only the probability is going to be reported.
Another limitation of these model checkers is that they cannot handle models with infinite state space. 
The methods developed for handling such models~\cite{roberts2022stamina} are not able to produce counterexamples as well. 

In non-probabilistic model checking, a counterexample is a single execution trace leading to an erroneous state.
In the probabilistic setting, considering a property of the form $P_{\leq p}(\Phi)$, a trace satisfying $\Phi$ is referred to as a \emph{witness}. 
Since counterexamples are very useful in helping the designer with locating the errors or for further refinements of the model, many methods have been proposed for the counterexample generation in stochastic model checking.
Han and Katoen\cite{han2007counterexamples} previously reduced the problem of counterexample generation in DTMCs to computing \emph{(hop-constrained) k-shortest paths} (KSP) in a graph. 
They further extended their work with a set of approximate methods to generate counterexamples for CTMC models\cite{han2007providing}. However, these methods do not scale to larger models.
Aljazzar and Leue\cite{aljazzar2009directed} proposed a variation of \emph{Best First Search} (BestFS) over k-shortest path in order to address scalability issues.
These methods only consider finite state Markov chains, and have not been applied to chemical reaction networks considered in this paper.
SAT-based bounded model checking is used to generate counterexample in~\cite{wimmer2009counterexample}, and it is extended in~\cite{braitling2011counterexample} by using SMT-based bounded model checking to generate smaller counterexamples and broaden its application to include Markov reward models as well.
These bounded methods are only applied to DTMCs but not CTMCs.
A counterexample for such property consists of a set of witnesses with their accumulated probability greater than $p$. A challenge in generating counterexamples for probabilistic models is that the set of traces needed for a counterexample can potentially grow very large. 

Many real-world systems cannot be modeled with bounded variables, leading to infinite-state Markov chain models. 
A particular type of such systems are \emph{chemical reaction networks} (CRNs). 
CRNs are a language used to describe biochemical systems~\cite{chellaboina2009modeling}.
A CRN describes the evolution of a biological system consisting of a set of species based on a set of chemical reaction rules. 
In order to analyze a CRN, its stochastic temporal behavior is first modelled as a CTMC, which is then analyzed by a probabilistic model checker.
Since there is usually no bound on the populations of chemical species in a realistic biological system, the corresponding CTMC model potentially has an infinite state-space.

Infinite state-space models pose some serious challenges to the previously established counterexample generation methods.
Any graph based algorithm that requires the whole state-space to be explicitly stored in memory would be useless. 
Therefore, $k$-shortest path method and its variations cannot be applied to infinite state models.
In the best first search algorithms such as \cite{aljazzar2009directed}, state-space can be expanded on the fly, and therefore it does not suffer the same issue as the KSP method. 
However, the BestFS algorithms are still not guaranteed to terminate on infinite state spaces, and the state space expanded in the BestFS algorithm can explode quickly.

This paper proposes a bounded model checking (BMC) based method to generate counterexamples for the CRN models  with potentially infinite state-space. 
In general, this method looks for a possible witness trace of length $k$.
If no such witnesses were found or the set of witnesses did not have enough probability already, the length $k$ is increased to search for more witnesses.
This process iterates until a counterexample is returned.
In this method, the counterexample generation for a CRN is first encoded as a constraint satisfaction problem.  
It is then solved with a SMT solver. 
If the problem is \emph{sat}, a witness can be extracted. 
Otherwise, no witness exists in that encoding.

The above simple idea does not require the state-space to be explicitly stored in memory, and unlike BestFS method, it is guaranteed to terminate if a counterexample exists. However, the performance of this method can suffer when a deep unrolling of the model is needed to find the very first witnesses or when a large number of witnesses needs to be found in order to refute a property.

To address the above problems, we propose two optimizations to improve the BMC method targeting CRNs.
The first optimization is to search for witness traces using a divide-and-conquer technique.
This technique limits the depth of unrolling required by BMC to generate witnesses, improving the performance of the method when a deep unrolling is needed to generate witnesses.
The second optimization is to construct new and longer witnesses using the current set of witnesses without the need of producing and solving large constraint problems.
It allows many new witnesses to be generated quickly, thus improving the overall performance of the proposed method when a large set of witnesses is required to refute the property.

This paper makes the following contributions.  
First, to our best knowledge, the counterexample generation framework developed in this paper is the first to target CRNs with infinite state space.
It extends previous BMC based methods to handle continuous-time Markov models.
\vspace{-1pt}
Second, novel optimizations are developed to tackle unique challenges facing the counterexample generation for CRNs.  
These optimizations enable long witnesses or a large number of witnesses  necessary for a counterexample to be generated efficiently without deep unrolling of CRN models, thus significantly reducing the complexity of the problems for the backend SMT solving engine.

\section{Background}
\label{sec:background}

This section defines chemical reaction networks and their semantics similar to those given in~\cite{crn-analysis:cav2019}, and then the counterexample generation problem.

This paper considers a subset of chemical reaction networks~(CRNs). 
The more general definitions for such networks can be found in~\cite{crn-analysis:cav2019}. 
A CRN consists of a finite number $N$ of chemical species, $\mathcal{S} = \{S_1, S_2, ..., S_N\}$, interacting with each other through a finite number $M$ of reaction channels $\mathcal{R} = \{R_1, R_2, ..., R_M\}$.
A \emph{reaction} $({Ra}, {Pd}, \lambda) \in \mathcal{R}$ defines a rule on the evolution of the system, 
where $Ra \subseteq \mathcal{S}$ is a set of species called $reactants$, $Pd \subseteq \mathcal{S}$ is a set of species called \emph{products}, and $\lambda$, which is a positive real number, is the coefficient associated with the rate of the reaction.
Note that $Ra$ and $Pd$ can be empty. 
A reaction is typically written as 
\begin{center}
$S_h + S_i + \ldots \xrightarrow[]{~\lambda~} S_j + S_k + \ldots$
\end{center}
given that $\{S_h, S_i, \ldots\} \subseteq Ra$ and $\{S_j, S_k, \ldots\} \subseteq Pd$.

The semantics of a CRN $\mathcal{C}$ is generally given in terms of a discrete-state continuous-time stochastic process. 
A state of a CRN, $\bm{x}(t)$, is a vector representing the populations of species $\mathcal{S}$ at time $t \geq 0$.  
A state without time explicitly represented is simply written as $\bm{x}$.
Firing of a reaction causes a state change by changing populations of some species. 
Let $\bm{x}[S_i]$ denote the population of species $S_i$ in state $\bm{x}$. 
A reaction $R = (Ra, Pd, \lambda)$ is \emph{enabled} in state $\bm{x}$ if $\forall S_i \in Ra, \bm{x}[S_i] > 0$.
Firing an enabled reaction in state $\bm{x}$ leads to a new state $\bm{x}^\prime$ such that the following conditions hold.
\[
\begin{array}{l}
     \forall S_i \in Ra - Pd,\ \bm{x}^\prime[S_i] = \bm{x}[S_i] - 1  \\
     \forall S_i \in Pd - Ra,\ \bm{x}^\prime[S_i] = \bm{x}[S_i] + 1 \\
     \forall S_i \in Pd \cap Ra,\ \bm{x}^\prime[S_i] = \bm{x}[S_i]
\end{array}
\]
The initial state of a CRN is denoted as $\bm{x}_0$. 

\begin{figure}[tb]
        \centering
        \includegraphics[height=1in]{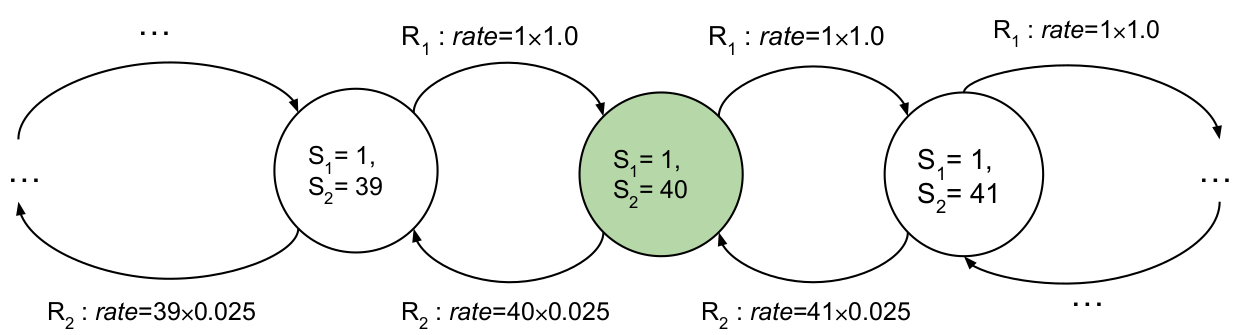}
        \caption{The single species production and degradation CRN modelled as a CTMC. The state in green is the initial state.}
        \label{fig:simple_crn}
\end{figure}

The stochastic behavior of a CRN $\mathcal{C}$ can be modeled as a possibly infinite continuous-time Markov chain (CTMC) $\mathcal{M} = (\bm{X}, \bm{x}_0, \bm{R})$ where $\bm{x}_0$ is the initial state of $\mathcal{C}$, $\bm{X}$ the set of reachable states from $\bm{x}_0$ via a sequence of reaction firings, and $\bm{R}$ the set of transitions.
Each transition in $\bm{R}$ is given by $\bm{x} \xrightarrow{r} \bm{x}'$ if there is a reaction $R = (Ra, Pd, \lambda)$ and a state $\bm{x}$ such that firing $R$ in state $\bm{x}$ leads to a new state $\bm{x}'$.
The transition rate $r$ is defined by a propensity function as follows.
\begin{equation}
\label{eq:propensity}
r = \lambda \times \prod_{\forall S_i \in Ra} \bm{x}[S_i].
\end{equation}
Note that the set of transitions $\bm{R}$ can also be expressed as a transition rate matrix as in the traditional CTMC definition.
As an example, the simple \emph{single species production and degradation} CRN taken from \cite{kuwahara2008efficient} is considered.
\begin{equation}
\label{eq:single-species}
\begin{array}{ll}
    R_1 : \ S_1 \xrightarrow{\lambda_1} S_1 + S_2,~~~ &
    R_2 : \ S_2 \xrightarrow{\lambda_2} \emptyset
\end{array}
\end{equation}
where the populations of species $S_1$ and $S_2$ in the initial state are $1$ and $40$ respectively,  and the reaction rate constants of the above two reactions are $\lambda_1 = 1.0$ and $\lambda_2 = 0.025$.
The resulting CTMC for that CRN model is shown in Fig.~\ref{fig:simple_crn}.

In this paper, the properties of interest specify that the probability of the system moving from the initial state to a state where population of species $S$ reaches $\theta$ within time limit $T$ is at most $p$ where $\theta$ is a natural number.
This type of properties can be formally defined as a CSL property of the form $P_{\leq p} [true\; U^{\leq T}\; S=\theta]$.  
For the CRN shown in Fig.~\ref{fig:simple_crn}, an example property is that the probability of the system moving from the initial state to a state where population of $S_2$ reaches $65$ within $100$ time units is at most $1\times10^{-4}$. 
It can be formally specified as the CSL property $P_{\leq 1\times10^{-4}} [true\; U^{\leq 100}\; S_2=65]$.

A trace starting from the initial state and ending in a state where $S=\theta$ is called a \emph{witness} to the property.
The goal of counterexample generation is to find the minimal set of such witnesses with accumulated probability greater than $p$.
\section{Counterexample Generation for CRNs}
\label{sec:method}

\subsection{Overview of the Framework}

Algorithm \ref{BMC_alg} shows the overview of the proposed counterexample generation framework using SMT-based BMC for CRN models.

The inputs to this framework are a CRN $\mathcal{C}$ and a property $P_{\leq p}[true \ U^{\leq T} S = \theta]$ for which a counterexample is to be generated.
The flow for the counterexample generation is as follows.
\begin{enumerate}
    \item First, starting with $k=0$, the CRN $\mathcal{C}$ is encoded as a BMC problem to generate a witness for property $S=\theta$ of bound $k$.  
Note that the witnesses found in this step have reaction rate information in the CRN~$\mathcal{C}$ abstracted away. A witness $w$ found in this step is a sequence of discrete states of $\mathcal{C}$, and has the following characteristics:

\begin{enumerate}[label=(\alph*)]
    \item The first state of $w$ is the initial state $\bm{x_0}$ of $\mathcal{C}$.
    \item $len(w)=k$, i.e. there are in total $k$ transitions in $w$.
    \item The last state in the witness, $\bm{x_k}$, satisfies the property $\bm{x_k}[S]=\theta$.
\end{enumerate}

    \item Next, the found witnesses are used to construct or expand a witness CTMC where transition rates in the CTMC are derived from reaction rates of $\mathcal{C}$ using the propensity function defined in~Eq.~(\ref{eq:propensity}).

    \item Once the witness CTMC is constructed, it is passed to a CTMC model checker, \emph{e.g.} PRISM, to compute its probability.  
    If the computed probability is higher than the $p$ specified in the input property $P_{\leq p}[true \ U^{\leq T} S = \theta]$, the flow terminates.
    Otherwise, the flow is repeated from step~1. If all the witnesses of bound $k$ are already generated, $k$ is incremented by $1$.
\end{enumerate}

\begin{algorithm}[tb]
\caption{Counterexample Generation for CRNs}
\begin{algorithmic} [1]
\State \textbf{Input:} CRN $\mathcal{C}$, Property $P_{\leq p}[true \ U^{\leq T} S = \theta]$
\State \textbf{Output:} $cex -$ counterexample refuting $P_{\leq p}[true \ U^{\leq T} S = \theta]$ for $\mathcal{C}$
\State $\Phi \gets S=\theta$
\State $witness\_set \leftarrow \emptyset$
\State $k \leftarrow 0$
\State $prob\leftarrow 0$
\State $encoding \leftarrow \mathit{BMC}(\mathcal{C}, k, \Phi)$
\While{$prob<p$}
    \If{${solve}(encoding) = \mbox{\emph{unsat}}$}
        \State $k \leftarrow k+1$
        \State $encoding \leftarrow \mathit{BMC}(\mathcal{C}, k, \Phi)$
    \Else
        \State $witness \leftarrow \mathit{create\_trace}(encoding)$
        \State $witness\_set \leftarrow witness\_set \cup witness$
        \State  $cex \leftarrow  \mathit{create\_ctmc}(witness\_set)$
        \State $encoding \leftarrow encoding \wedge {exclude}(witness)$
        \State $prob \leftarrow \mathit{calculate\_probability}(cex)$
    \EndIf
\EndWhile
\State \Return (cex) 
\end{algorithmic}
\label{BMC_alg}
\end{algorithm}

Solving the BMC encoding for a bound $k$ will result in either \emph{unsat} or a satisfying set of assignments are returned by the solver.
If the result is \emph{unsat}, either no witness of length $k$ exists in the model or all such witnesses are already found.
In this case, $k$ is incremented by one and a new encoding is generated to find longer witnesses.
If the solver returns \emph{sat}, the satisfying set of assignments are turned into a witness.
In order to prevent the solver from finding the same witness again, the BMC encoding is updated with additional constraints to exclude the already found witnesses.

Unlike DTMCs, where the probability of a trace is the product of the probability of the transitions along that trace, for CTMCs there is no such straightforward manner to calculate the probability of a trace and therefore a CTMC model checker needs to be utilized.
In general, calculating the probability of the set of witnesses can be done by turning every witness into a CTMC and calling a model checker to calculate the probability of that witness and then summing those probabilities.
The set of witnesses required to refute a property can be quite large, and calculating the probability for every single witness could result in a large number of calls to the CTMC model checker.
Therefore we propose a different approach to limit the number of calls to a model checker.
Instead of calculating the probability of every single witness, we first generate a \emph{witness \mbox{CTMC} graph} constructed by overlaying all the witnesses on top of each other.
Next, its probability is calculated by a CTMC model checker.
If this probability surpasses the threshold defined in the property, the witness CTMC graph is returned as the counterexample.
By performing this procedure whenever the size of the witness set increases by a predefined value, we can manage to limit the number of times a CTMC model checker is called and therefore limit the overhead it imposes on the framework.

\subsection{BMC Encoding}
\label{sec:bmc-encoding}

In \cite{wimmer2009counterexample}, Wimmer et al. proposed a method to turn the problem of finding a witness trace of certain length in a DTMC model into a propositional satisfiability (SAT) problem. 
Braitling et al. \cite{braitling2011counterexample} further extended this and proposed a method to turn the problem of finding a witness path of certain length in a DTMC model into a satisfiability modulo theories (SMT) problem that enforces a minimum threshold for the probability of found witnesses.
For a certain bound $k$, their extension allows to find witnesses of length $k$ with higher probability first, potentially decreasing the size of the counterexample set.

Here we propose a similar approach to turn the problem of finding a witness of a fixed length in a CRN $\mathcal{C}$ into a SMT solving problem.
Consider a CRN characterized by $N$ species $\mathcal{S}=\{S_1, S_2, ..., S_N\}$ and $M$ reactions $\mathcal{R} = \{R_1, R_2, ..., R_M\}$.
The proposed method first encodes the CRN as follows.
The initial state $\bm{x}_0$ is encoded as 
\[
enc_0(\bm{x}_0) := \bigwedge_{i = 1,\ldots, N} v^0_i = x_0[S_i].
\]
where $v^0_i$ are integer variables, one for each species $S_i$.
For $k>0$, each reaction $R_i: Ra \xrightarrow{\lambda} Pd$ is encoded as
\[
enc(R_i, k) := \left(\bigwedge_{S_i \in Ra} v^{k-1}_i >0 \right) \wedge \left( \alpha \wedge
      \beta \wedge 
      \gamma \right)
\]
\[
\alpha := \bigwedge_{\forall S_i \in Ra - Pd} v^{k-1}_i - 1 = v^{k}_i
\]
\[
\beta:= \bigwedge_{\forall S_i \in Pd - Ra} v^{k-1}_i + 1 = v^{k}_i
\]
\[
\gamma:= \bigwedge_{\forall S_i \in Pd \cap Ra} v^{k-1}_i = v^{k}_i
\]
where $v^k_i$ are integer variables parameterized by bound $k$, one for each species $S_i$.
We abuse the notation $S_i \notin R_i$ to indicate that species $S_i$ is not present in reaction $R_i$.
Then, the entire set of reactions $R$ is encoded as follows.
\[
enc(R, k) := \bigvee_{R_i \in R} \left(enc(R_i, k) \wedge (\bigwedge_{\forall{S_i \notin R_i}} v^{k-1}_i = v^{k}_i)\right)
\]
Next, the constraint for loops of a witness is encoded as follows. 

\begin{equation}
\label{eq:loop-enc}
loop(k) := \bigvee_{0 \leq h \leq k}\; \bigvee_{0 \leq j < h}\; \bigwedge_{\forall S_i \in \mathcal{S}} v^h_i = v^j_i.
\end{equation}
The above constraint holds true for a witness $x_0, \ldots, x_j, \ldots, x_h, \ldots, x_k$ such that $x_h$ is a transition back to $x_j$ by some reaction.
In BMC~\cite{biere1999symbolic}, such transitions are referred to as \emph{back edges}.

Since the BMC aims to find traces of a CRN model that reach the target states that satisfy $\phi := S_i=\theta$ for a species $S_i$, the target states are encoded as 
\[
enc_t(\phi, k) := v^k_i = \theta.
\]
Combining the above encodings, the problem of finding traces that reach the target states $S_i = \theta$ in $k$ steps is encoded as a constraint satisfaction problem
\begin{equation}
\label{eq:bmc-encoding}
 BMC(\mathcal{C}, \phi, k) := enc_0(x_0) \wedge \bigwedge_{0\leq i\leq k} enc(R, i) \wedge \neg{loop(k)} \wedge enc_t(\phi, k).   
\end{equation}
Subsequently, the above constraint encoding can be fed into an SMT solver.  
If the solver finds it satisfiable, a witnessing trace of length $k$ can be returned.  

The above encoding avoids finding loop induced witnesses for more efficient SMT solving. 
On the other hand, presence of loops on a witness can increase the probability of such witness, therefore the loops need to be recovered.  
In our method, after a loop-free witness is found, back edges constituting loops are found with the scaffolding method, described later in this section.
Then, that witness and the found back edges are used to construct counterexample CTMC as described in the next section.

For a certain $k$, if (\ref{eq:bmc-encoding}) returns \emph{unsat}, then there is no witness of length $k$ starting from the initial state and ending in a state satisfying the property. 
If (\ref{eq:bmc-encoding}) returns \emph{sat}, the satisfactory assignments to variables characterize a witness to the property.
For a given bound $k$, if (\ref{eq:bmc-encoding}) returns \emph{sat}, solving it again would result in the same assignments characterizing the same witness. In order to prevent the solver from generating the same witness, a new constraint explicitly stating to exclude already found assignments from the set of possible solutions is added to the SMT formulation.

\subsection{Witness CTMC Construction}
\label{sec:ctmc_construction}

As mentioned before, calculating the probability of a witness for a CTMC model requires utilizing a model checker.
If the set of witnesses required to refute a property is large, calculating the probability of each individual witness requires multiple calls to a model checker and this would impact the performance of the framework.
Instead, a \emph{witness}~CTMC graph is generated from a set of witnesses, and then passed to a CTMC model checker for finding the probability.

The witness CTMC is constructed as follows.
For each individual witness $w$, for each state $\bm{x}$ of $w$, if it is not already in the witness CTMC, a new vertex corresponding to $\bm{x}$ is added.  
For each transition $\bm{x} \xrightarrow[]{R} \bm{x}'$ in $w$, an edge corresponding to that transition is added between two vertices corresponding to $\bm{x}$ and $\bm{x}'$ of that transition.
The transition rate of that edge is calculated by the propensity function as defined in Eq.~(\ref{eq:propensity}).
Given a new witness, the witness CTMC can be expanded readily as described above.
Additionally, in order to maintain the semantics of the CRN model, in every state of the witness CTMC, if a reaction of the CRN model is enabled in that state, but absent from the witnesses, a transition to a unique \emph{sink} state is added correspondingly, with the transition rate computed using the propensity function defined in Eq.~(\ref{eq:propensity}).

\begin{figure}[tb]
    \centering
\begin{tabular}{c}
    \includegraphics[width=0.7\textwidth]{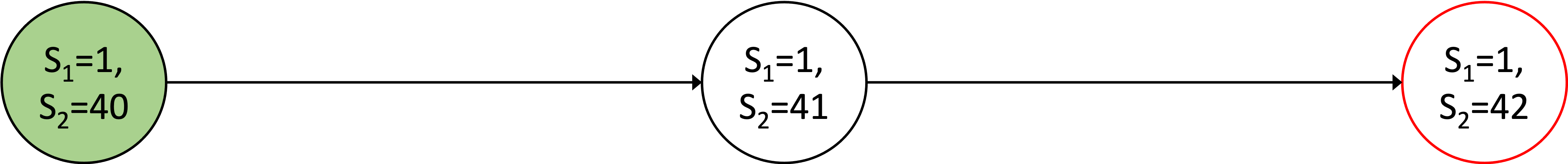}\\
    (a) 
    \\[9pt]

    \includegraphics[width=0.7\textwidth]{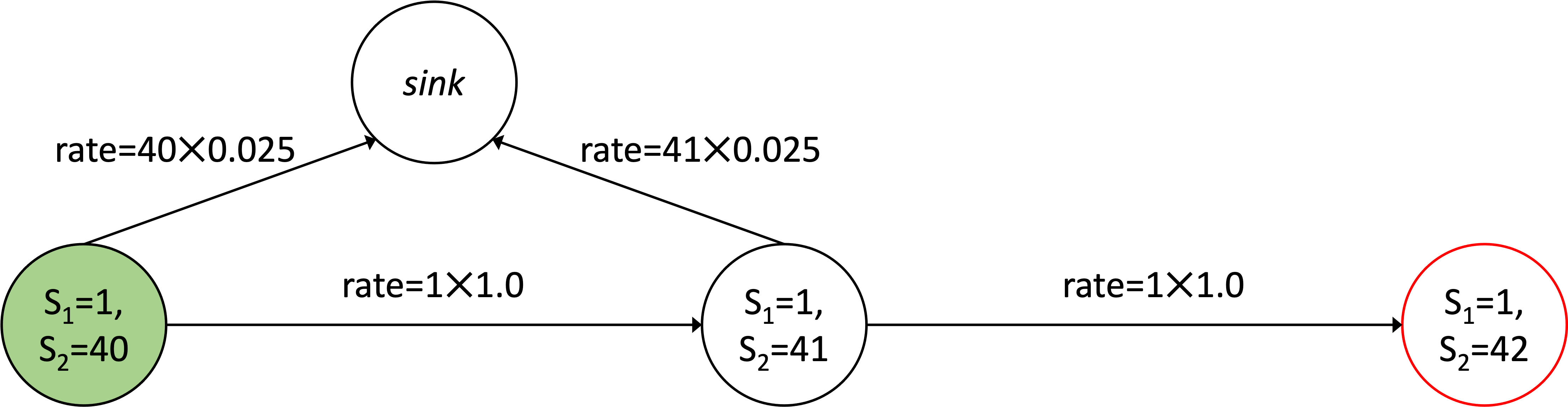} \\
    (b)
    \\[9pt]

    \includegraphics[width=0.7\textwidth]{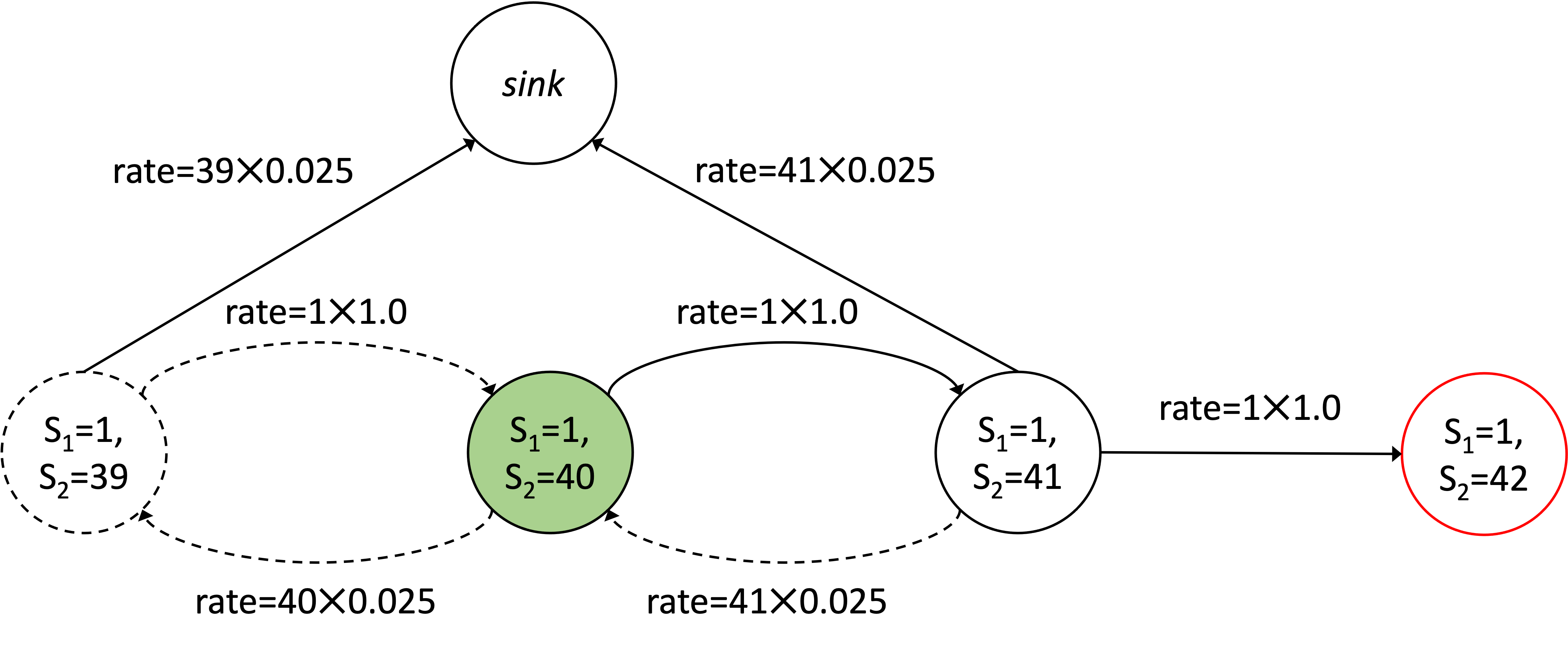} \\
    (c)

\end{tabular}
\caption{(a) A witness generated for single species production and degradation model with bound $k=2$ drawn as a graph, (b) The CTMC constructed from that witness, (c) The witness CTMC expanded using the scaffolding method with bounds $j \leq 2$. New states and transitions found are dashed.}       
\label{fig:construct_and_scaffold}
\end{figure}

As an example, consider the simple model depicted in Fig. \ref{fig:simple_crn} and the property specifying target states $S_2=42$. 
First a witness to this property is found at bound $2$, and it is shown in Fig.~\ref{fig:construct_and_scaffold}(a). 
Fig. \ref{fig:construct_and_scaffold}(b) shows the resulting CTMC constructed from the witness shown in Fig. \ref{fig:construct_and_scaffold}(a) where edges are labeled with transition rates derived from species' populations and reaction rate constants using the propensity function in Eq.~(\ref{eq:propensity}).
Also note the additional edges going to the sink state to preserve the stochastic behavior of the CRN model. 

Generating a witness CTMC graph instead of calculating the probability of every single witness allows the framework to call a CTMC model checker whenever the size of the witness CTMC graph has increased by a predefined threshold, significantly reducing the number of calls to the model checker.

\subsection{Optimizations}

\subsubsection{Divide-and-Conquer}

One particular challenge in using BMC for counterexample generation for CRNs appears when a model unrolling with a large bound is required to generate first witnesses.
Suppose that the first witness appears at bound $k$.
BMC first goes through bounds $0,1,...,k-1$, and refutes the existence of witnesses at those bounds.
If $k$ is relatively large, it causes two issues.
First, the above refuting process can be computationally expensive.
Second, the BMC encoding for a large bound $k$ makes the SMT solving difficult.

This problem is exacerbated for CRN models as they typically have dense state spaces, \ie, those where a large number of reactions are enabled in every state.
Therefore, the search space for witnesses can grow exponentially as the bound $k$ increases.

In order to alleviate the problem, we propose a divide-and-conquer approach exploiting the structure of the CRNs.
Note that for a CRN $\mathcal{C}$ and a property $P_{\leq p}[true \; U^{\leq T} S_i=\theta]$, firing a reaction causes one of possible changes to the population of $S_i$: incremented by $1$, decremented by $1$, or remained unchanged.
Now suppose that $\bm{x}_0[S_i]=\eta$.
We first select a positive number $\delta$ such that $(\theta-\eta)\%\delta=0$.
Let $n = (\theta-\eta)/\delta$.
Next, BMC is used to find a trace segment $\bm{x}_0, \ldots, \bm{x}_{\delta}$ such that $\bm{x}_{\delta}[S_i]=\eta + \delta$.
Then, a second trace segment $\bm{x}_{\delta}, \ldots, \bm{x}_{2\delta}$ is found such that $\bm{x}_{2\delta}[S_i]=\eta + 2\delta$.
This procedure continues until a trace segment $\bm{x}_{(n-1)\delta}, \ldots, \bm{x}_{n\delta}$ is found such that $\bm{x}_{(n-1)\delta}[S_i]=\eta+(n-1)\delta$ and $\bm{x}_{n\delta}[S_i]=\eta+n\delta$.
If all such trace segments are found, they are concatenated to form a witness.

As an example, consider the CRN model given in Fig. \ref{fig:simple_crn} and a property where $\theta=70$.
Using the divide-and-conquer approach with $\delta=10$, three trace segments are found. 
To find each trace segment, the CRN model is just unrolled for $10$ steps for the BMC encoding. 
In comparison, the original BMC encoding needs to unroll the CRN model for $30$ steps in order to find witnesses, thus causing a more complex problem to solve.

Note that the above technique may not be able to find all $n$ trace segments necessary to form a witness. 
In that case, this technique is simply terminated, and the framework rolls back to the original flow.

\subsubsection{Scaffolding}

The other challenge facing the counterexample generation for CRNs is that a large number of witnesses need to be found.  
Recall that a witness CTMC is constructed from the witnesses as a candidate counterexample.  
It becomes a counterexample if it accumulates enough probability.
Otherwise, it is expanded with additional witnesses. 

Depending on the model and the property, the size of the witness CTMC required to refute the property might be very large.
This is especially true if the probability of a property is not concentrated in a relatively small region of the model's state space, but distributed among many witnesses, each of which contributes only a small amount of probability to the final counterexample.
In such cases, the SMT solver needs to be repeatedly invoked for finding new witnesses, which is computationally expensive.
This problem is significantly exacerbated when the BMC encoding becomes large with large bounds $k$.
Therefore, expanding the witness CTMC by adding many witnesses can be very slow.
A \emph{scaffolding} method is described to mitigate the above challenge.

When a witness is found, it is guaranteed that the target state is reachable from every state of that witness. 
Therefore, a trace sharing a suffix of a witness is also a witness.
The scaffolding method exploits the existing witnesses, instead of solving a BMC encoding of a large bound $k$, to find many new witnesses quickly. The procedure is as follows.
Suppose $\bm{W}$ is a non-empty set of witnesses, and $\bm{\tau}$ the target state.
Let $\bm{X_W}$ be the set of all the states present in $\bm{W}$.
New witnesses can be constructed by finding traces of length $j$ reaching $\bm{X_W} \cup \bm{\tau}$ from $\bm{X_W}$. 
Note that the scaffolding method is very similar to the BMC method for counterexample witness generation presented in Section~\ref{sec:bmc-encoding} except for the encodings for initial and target states.
The BMC encoding for the scaffolding method is given below.
\begin{center}
    $BMC(\mathcal{C}, \bm{W}, j) := enc(\bm{X_W}, 0) \wedge \bigwedge_{0\leq i\leq j} enc(R, i) \wedge enc(\bm{X_W} \cup \bm{\tau}, j). $  
\end{center}
where 
\begin{center}

$
enc(\bm{X_W},0) = \bigvee_{\bm{x} \in \bm{X_W}} enc_0(\bm{x}) \mbox{~~and~~} 
enc(\bm{X_W} \cup \bm{\tau}, j) = \bigvee_{\bm{x} \in \bm{X_W} \cup \bm{\tau}} enc_t(\bm{x}, j).
$
\end{center}
The new witnesses found by the scaffolding method on the existing witnesses of bound $k$ have lengths of at least $k+j$ and are generated by solving a BMC encoding for bound $j$.

As an example, consider the model given in Fig.~\ref{fig:simple_crn} and the property specifying the target state with $S_2=42$.
After the first witness as shown in Fig.~\ref{fig:construct_and_scaffold}(a) is generated with bound $2$ using the original BMC encoding as in~(\ref{eq:bmc-encoding}), the scaffolding method solves a new BMC encoding with bound $j=2$, and generates new witnesses, which lead to one new state $[S_1=1, S_2=39]$ and three new edges added to the witness CTMC in Fig.\ref{fig:construct_and_scaffold}(b).

The expanded CTMC is shown in Fig.~\ref{fig:construct_and_scaffold}(c).
Note that we need to solve an encoding of bound $4$ using the original BMC method to find the witness with the new state and transitions.
However, the scaffolding method finds it by solving a smaller encoding of bound $2$.
Therefore, it can significantly increase the speed of witness generation.

\section{Experimental Results}

In this section we report the results obtained by testing the proposed framework on four infinite-state CRNs.
The described framework is implemented in Python with Z3~\cite{moura2008z3} as the underlying SMT-solver.
The experiments are performed on an Ubuntu virtual machine running on a PC with a $3.6$GHz Intel processor.
The virtual machine is allocated three cores and $8$GB of main memory.
Any experiment that took more than $1800$ seconds was terminated and is marked with -TO- in the result tables. 
Any experiment that used more than $2$GB of main memory was also aborted, and is marked with -MO-. 
We compare our framework against \emph{DiPro}~\cite{aljazzar2011dipro}, which is the Java implementation of the extended best first search algorithm described in \cite{aljazzar2009directed}.
In all of the following experiments scaffolding method is used with the following parameters:
After finding every $3$ witnesses to the property using either original BMC flow or divide-and-conquer approach, scaffolding method is called with $j\leq3$ to generate $50$ new witnesses from the set of currently found witnesses. 

In order to obtain a lower probability bound for the properties of interest, we limit the range of values the variables can take in order to be able to use a model checker. However, the counterexample generation method is run using the model with unbounded variables in all experiments.

\subsection{Single species production-degradation model}

The first model being evaluated is the single species production and degradation as shown in~$\eqref{eq:single-species}$.
The initial populations of $S_1$ and $S_2$ are $1$ and $40$ respectively.

Bounding variables $S_1$ and $S_2$ to the range $[0,70]$, PRISM reports that the probability of the system moving from initial state to a state where $S_2$ population equals $70$ within time limit $100$ is $1.67\times10^{-4}$.
Therefore, for the CSL property $P_{\leq p}[\textrm{true}\; U^{\leq100}\; S_2=70]$ a counterexample can be found for any $p<1.67\times10^{-4}$.
The divide-and-conquer technique is not used for this simple model.

Table~\ref{tab:single} shows the results for running the described BMC framework and DiPro to generate counterexamples for $3$ different threshold values. The size of a counterexample is defined as the sum of the number of states and the number of transitions in the counterexample CTMC.
DiPro expands the witness CTMC much more quickly and finds the portion of state space containing all the probability for the property regardless of the probability threshold.
BMC expands the witness CTMC slower, but the returned counterexample is smaller, specially for lower thresholds.

\begin{table}[tb]
    \caption{Counterexamples for checking the single species model against CSL property  $P_{\leq p}[\textrm{true}\; U^{\leq100}\; S_2=70]$. First column shows the value for $p$ in the property. Time is measured in seconds.}   
    \begingroup
    \renewcommand{\arraystretch}{1.25} 
    \[
    \begin{tabular}{|c|c|c|c||c|c|c|}
        \hline
        \multicolumn{1}{|c|}{} & \multicolumn{3}{|c||}{\textbf{BMC}} & \multicolumn{3}{|c|}{\textbf{DiPro}} \\
        \hline
        $\mathbf{p}$ & \textbf{cex prob.} & \textbf{time} & \textbf{cex size} & \textbf{cex prob.} & \textbf{time} & \textbf{cex size} \\
        \hline
         $1\times10^{-20}$ & $7.14\times10^{-12}$ & $0.6$ & $61$ & $1.67\times10^{-4}$ & $0.07$ & $168$\\
         $1\times10^{-10}$ & $7.04\times10^{-5}$ & $2.5$ & $94$ & $1.67\times10^{-4}$ & $0.07$ & $168$\\
         $1\times10^{-4}$ & $1.16\times10^{-4}$ & $5.4$ & $100$ & $1.67\times10^{-4}$ & $0.07$ & $168$\\
         $1.5\times10^{-4}$ & $1.52\times10^{-4}$ & $9.8$ & $121$ & $1.67\times10^{-4}$ & $0.07$ & $168$\\
         \hline
    \end{tabular}
    \]
    \endgroup
    \label{tab:single}
\end{table}

\subsection{Enzymatic futile cycle model}

Next, the \emph{Enzymatic Futile Cycle} model taken from \cite{kuwahara2008efficient} is considered. This model is defined by the following $6$ reactions:
\[
\begin{array}{lll}
    R_1 : \ \textrm{S}_1 + \textrm{S}_2 \xrightarrow{1.0} \textrm{S}_3,~~~ &
    R_2 : \ \textrm{S}_3 \xrightarrow{1.0} \textrm{S}_1 + \textrm{S}_2,~~~ &
    R_3 : \ \textrm{S}_3 \xrightarrow{0.1} \textrm{S}_1 + \textrm{S}_5, \\
    R_4 : \ \textrm{S}_4 + \textrm{S}_5 \xrightarrow{1.0} \textrm{S}_6,~~~ &
    R_5 : \ \textrm{S}_6 \xrightarrow{1.0} \textrm{S}_4 + \textrm{S}_5,~~~ &
    R_6 : \ \textrm{S}_6 \xrightarrow{0.1} \textrm{S}_4 + \textrm{S}_2
\end{array}
\]
where the initial populations of species $(S_1, S_2, S_3, S_4, S_5, S_6)$ are 
\begin{center}
$\bm{x_0} = [1, 50, 0, 1, 50, 0].$
\end{center}

Bounding all the species' population to be in the range $[0,100]$, The probability that the system moves to a state where population of $S_5$ is $40$ within $100$ time units is calculated to be $0.042$ by PRISM.
Therefore, a counterexample can be generated for the CSL property $P_{\leq p}[true \; U^{\leq 100} S_5=40]$ if $p$ is set to any value less than $0.042$.
Since generating witnesses does not need a deep unrolling of this model and since the state space does not have high density, the divide-and-conquer approach does not make any significant yield in performance and therefore is not used in this experiment. 

\begin{table}[tb]
    \caption{Counterexamples for checking the enzymatic futile cycle model against CSL property  $P_{\leq p}[\textrm{true}\; U^{\leq100}\; S_5=40]$. First column shows the value for $p$ in the property. Time is measured in seconds.}   
    \begingroup
    \renewcommand{\arraystretch}{1.25} 
    \[
    \begin{tabular}{|c|c|c|c||c|c|c|}
        \hline
        \multicolumn{1}{|c|}{} & \multicolumn{3}{|c||}{\textbf{BMC}} & \multicolumn{3}{|c|}{\textbf{DiPro}} \\
        \hline
        \textbf{p} & \textbf{cex prob.} & \textbf{time} & \textbf{cex size}  & \textbf{cex prob.} & \textbf{time} & \textbf{cex size} \\
        \hline
         $1\times10^{-30}$ & $8.04\times10^{-29}$ & $2.8$ & $39$ & $3.1\times10^{-2}$ & $0.09$ & $260$\\
         $1\times10^{-20}$ & $2.87\times10^{-2}$ & $13.2$& $158$ & $3.1\times10^{-2}$ & $0.09$ & $260$\\
         $4\times10^{-2}$ & $4.13\times10^{-2}$ & $29.2$ & $190$ & $4.2\times10^{-2}$ & $0.16$ & $295$ \\
        \hline
    
    \end{tabular}
    \]
    \endgroup
    \label{tab:enzym}
\end{table}

Table~\ref{tab:enzym} shows the results for running the described BMC framework and DiPro to generate counterexamples for $3$ different threshold values.
Again, DiPro expands the witness CTMC much more quickly and is less sensitive to the probability thresholds in the property.
The proposed method is slower than DiPro for this model, but the counterexamples generated are smaller and hence have higher quality.

\subsection{Yeast polarization}

The \emph{Modified Yeast Polarization} model, taken from \cite{daigle2011automated}, is a CRN consisting of $7$ species reacting through $8$ reaction channels.
\[
\resizebox{\textwidth}{!}{
$\displaystyle
\begin{array}{lll}
    R_1 : \ \emptyset \xrightarrow{0.0038} \textrm{R}, &
    R_2 : \ \textrm{R} \xrightarrow{4.00\times 10^{-4}} \emptyset,~~~ &
    R_3 : \ \textrm{L} + \textrm{R} \xrightarrow{0.042} \textrm{RL} + \textrm{L}, \\
    R_4 : \ \textrm{RL} \xrightarrow{0.0100} \textrm{R},~~~ &
    R_5 : \ \textrm{RL} + \textrm{G} \xrightarrow{0.011} \textrm{G}_\textrm{a} + \textrm{G}_{\textrm{bg}},~~~ &
    R_6 : \ \textrm{G}_\textrm{a} \xrightarrow{0.100} \textrm{G}_\textrm{d}, \\
    R_7 : \ \textrm{G}_\textrm{d} + \textrm{G}_{\textrm{bg}} \xrightarrow{1.05\times 10^{3}} \textrm{G},~~~ &
    R_8 : \ \emptyset \xrightarrow{3.21} \textrm{RL} & \\
\end{array}$
}
\]
where the initial populations of species 
$(R, L, RL, G, G_{a}, G_{bg}, G_d)$
are 
\begin{center}
$\bm{x_0} = [50, 2, 0, 50, 0, 0, 0].$
\end{center}

The property of interest for this model is the population of $G_{bg}$ reaching $50$ within $20$ time units.
Bounding the species' population to the range $[0,150]$, STAMINA~\cite{roberts2022stamina} reports the lower bound of $1.64\times10^{-6}$ for this property.
Therefore, the CSL property $P_{\leq p}[true \; U^{\leq20} \; G_{bg}=50]$ is going to be refuted for any $p<1.64\times10^{-6}$ and a counterexample can be generated.

This model shows a very dense state space with many reactions enabled in a big portion of state space.
Also by observing the structure of the model, it can be deduced that the shortest witness to the property has $100$ transitions and therefore a deep unrolling of BMC is required to generate witnesses.

Without using the divide-and-conquer technique, we are not able to generate any witnesses within $1800$ seconds of running the program. 
Therefore, the divide-and-conquer method is used with the step set to $5$.
This means that in order to find a witness from the initial state to a state where $G_{bg}=50$ we first generate a trace from the initial state to a state where $G_{bg}=5$.
Then using this newly found state as the initial state, we find a witness to a state where $G_{bg}=10$ and continue this procedure until we reach a final state where $G_{bg}=50$.

The results obtained by running the program with the described framework are given in Table \ref{tab:enzym}. Note that DiPro could not generate a counterexample for any of the $5$ thresholds within the memory limitations. BMC framework could not generate a counterexample for thresholds greater than or equal to $1\times10^{-50}$ within the preset time limits.
\begin{table}[tb]
    \caption{Counterexamples for checking the modified yeast polarization model against CSL property  $P_{\leq p}[\textrm{true}\; U^{\leq20}\; G_{bg}=50]$. First column shows the threshold value $p$ in the property. Time is measured in seconds.}   
    \begingroup
    \renewcommand{\arraystretch}{1.25} 
    \[
    \begin{tabular}{|c|c|c|c||c|c|c|}
    \hline
        \multicolumn{1}{|c|}{} & \multicolumn{3}{|c||}{\textbf{BMC}} & \multicolumn{3}{|c|}{\textbf{DiPro}} \\
        \hline
        \textbf{p} & \textbf{cex prob.} & \textbf{time} & \textbf{cex size} & \textbf{cex prob.} & \textbf{time} & \textbf{cex size} \\
        \hline
         $1\times10^{-90}$ & $2.54\times10^{-90}$ & $9.4$ & $333$ & -MO- & - & -\\
         $1\times10^{-80}$ & $1.08\times10^{-80}$ & $25.2$ & $681$ & -MO- & - & -\\
         $1\times10^{-70}$ & $1.09\times10^{-70}$ & $122.7$ & $1620$ & -MO- & - & -\\
         $1\times10^{-60}$ & $1.11\times10^{-60}$ & $1207.6$ & $4593$ & -MO- & - & -\\
         $1\times10^{-50}$ & -TO- & - & - & -MO- & - & - \\
         \hline
    \end{tabular}
    \]
    \endgroup
    \label{tab:yeast}
\end{table}

\subsection{Motility regulation}
For this case study we take \emph{Motility Regulation} model from \cite{gillespie2019guided},  which is a CRN with a dense state space consisting of $9$ species reacting through $12$ reaction channels.
This model is defined by the following set of reactions:
\[
\resizebox{\textwidth}{!}{
$\displaystyle
\begin{array}{ll}
    R_1 : \ \textrm{codY} \xrightarrow{0.1} \textrm{codY} + \textrm{CodY}, &
    R_2 : \ \textrm{CodY} \xrightarrow{0.0002} \emptyset, \\
    R_3 : \ \textrm{flache} \xrightarrow{1} \textrm{flache} + \textrm{SigD}, &
    R_4 : \ \textrm{SigD} \xrightarrow{0.0002} \emptyset, \\
    R_5 : \ \textrm{SigD\_hag} \xrightarrow{1} \textrm{SigD} + \textrm{hag} + \textrm{Hag}, &
    R_6 : \ \textrm{Hag} \xrightarrow{0.0002} \emptyset, \\
    R_7 : \ \textrm{SigD} + \textrm{hag} \xrightarrow{0.01} \textrm{SigD\_hag}, &
    R_8 : \ \textrm{SigD\_hag} \xrightarrow{0.1} \textrm{SigD} + \textrm{hag}, \\
    R_9 : \ \textrm{CodY} + \textrm{flache} \xrightarrow{0.02} \textrm{CodY\_flache},~~ &
    R_{10} : \ \textrm{CodY\_flache} \xrightarrow{0.1} \textrm{CodY} + \textrm{flache}, \\
    R_{11} : \ \textrm{CodY} + \textrm{hag} \xrightarrow{0.01} \textrm{CodY\_hag}, &
    R_{12} : \ \textrm{CodY\_hag} \xrightarrow{0.1} \textrm{CodY} + \textrm{hag} \\
\end{array}$
}
\]
where the initial populations of the species 
\begin{center}
$(\mbox{codY}, \textrm{CodY}, \textrm{flache}, \textrm{SigD}, \textrm{SigD\_hag}, \textrm{hag}, \textrm{Hag}, \textrm{CodY\_flache}, \textrm{CodY\_hag})$
\end{center}
are $\bm{x_0} = [1, 10, 1, 10, 1, 1, 10, 1, 1].$

For this model we are checking for the property that $CodY$ population reaches $19$ within $10$ time units starting from the initial state.
Bounding the species' population to the range $[0,100]$, PRISM reports that the probability of this property is $2.49\times10^{-6}$.
Therefore, a counterexample can be generated for the CSL property $P_{\leq p}[true \; U^{\leq 10} CodY=19]$ for any $p<2.49\times10^{-6}$.

For this model we utilize the divide-and-conquer method with the step set to $3$.
This means that in order to find a witness to a state where $CodY=19$ we first generate a trace from the initial state to a state where $CodY=13$.
Then using this newly found state as the initial state, we find a witness to a state where $CodY=16$ and using this state as the new initial state, finally a trace to the target state with $CodY=19$ is found.

The results obtained by running the program with the described framework are given in Table \ref{tab:motil}. DiPro crashed on this model throwing an index out of bound error, and only the results for running the proposed BMC framework are reported. Proposed BMC framework is able to generate a counterexample for thresholds up to values very close to the total probability of the property.
    
\begin{table}
    \caption{Counterexamples for checking the motility regulation model against CSL property  $P_{\leq p}[\textrm{true}\; U^{\leq10}\; CodY=19]$. First column shows the value for $p$ in the property. Time is measured in seconds.}   
    \begingroup
    \renewcommand{\arraystretch}{1.25} 
    \[
    \begin{tabular}{|c|c|c|c|c|c|}
        \hline
        \multicolumn{1}{|c|}{} & \multicolumn{3}{|c|}{\textbf{BMC}}\\
        \hline
        \textbf{p} & \textbf{cex prob.} & \textbf{time} & \textbf{cex size} \\
        \hline
         $1\times10^{-20}$ & $3.38\times10^{-15}$ & $0.48$ & $19$\\
         $1\times10^{-10}$ & $6.49\times10^{-7}$ & $3.3$ & $1913$\\
         $1\times10^{-6}$ & $1.01\times10^{-6}$ & $197.8$ & $2876$\\
         $2\times10^{-6}$ & $2.04\times10^{-6}$ & $1483.4$ & $7340$\\
         \hline
    \end{tabular}
    \]
    \endgroup
     \label{tab:motil}
\end{table}

\subsection{Observations and Discussion}

For the first two experiments, both BMC-based method and DiPro are able to produce counterexamples for thresholds up to values very close to the total probability of the corresponding properties. 
In both experiments, DiPro builds the witness CTMC quickly to include almost the whole state space of the CRN models where witnesses to the property reside, and terminates much faster than the BMC approach. 
Our BMC based approach produces much smaller counterexamples  that are more comprehensible and useful for debugging purposes.

For the yeast polarization model, neither DiPro nor the original BMC approach without the divide-and-conquer optimization can produce a counterexample for any of the threshold values within the memory/time limits. 
Utilizing divide-and-conquer, our BMC-based approach is able to produce counterexamples for most of those thresholds.
The BMC-based approach is not able to find a counterexample for a larger threshold. 
Our understanding is that each witness carries only a tiny amount of probability, therefore many witnesses need to be found to accumulate sufficient probability. 
This can be very expensive in terms of runtime.
However, DiPro cannot handle this model at all.

For the motility regulation model, BMC framework using divide-and-conquer approach is able to generate counterexamples for threshold values close to the total probability of the property. 
Without using divide-and-conquer, the BMC-based approach is not able to generate counterexample for $p=2 \times 10^{-6}$ within the given time limit.
This and the previous experiments show the importance of that divide-and-conquer optimization.

\section{Conclusion}
We describe a BMC-based method 
to address counterexample generation problem for CRNs with potentially infinite state-space. 
Two optimizations are also proposed in order to scale it for larger and more complex models. 
In the experiments, we observe that the divide-and-conquer optimization is critical for larger and more complex CRNs where the counterexample witnesses can only be found by deep unrolling of the models with large bounds.
Additionally, the scaffolding method is developed to accelerate the counterexample generation without suffering the complexity of solving large BMC problems. 

Our framework does not order witnesses with respect to their probabilities.
We plan to further extend the framework to have a measure for probability of potential witnesses and find witnesses with higher probability first. \cite{braitling2011counterexample} proposes a method to implement this for DTMCs. We plan extend that to CRNs/CTMCs. 
We expect that such method, combined with techniques described in this paper will significantly improve the performance of the framework.

\paragraph{\bf Acknowledgements}
The authors are supported by the National Science Foundation under Grant Nos. 1856733, 1856740, and 1900542. Any opinions, findings, and conclusions or recommendations expressed in this material are those of the author(s) and do not necessarily reflect the views of the funding agencies.

%
%
%
\bibliographystyle{splncs04}
%
\bibliography{ref}{}

\begin{thebibliography}{10}
\providecommand{\url}[1]{\texttt{#1}}
\providecommand{\urlprefix}{URL }
\providecommand{\doi}[1]{https://doi.org/#1}

\bibitem{aljazzar2011dipro}
Aljazzar, H., Leitner-Fischer, F., Leue, S., Simeonov, D.: Dipro-a tool for
  probabilistic counterexample generation. In: International SPIN Workshop on
  Model Checking of Software. pp. 183--187. Springer (2011)

\bibitem{aljazzar2009directed}
Aljazzar, H., Leue, S.: Directed explicit state-space search in the generation
  of counterexamples for stochastic model checking. IEEE Transactions on
  Software Engineering  \textbf{36}(1),  37--60 (2009)

\bibitem{aziz1996verifying}
Aziz, A., Sanwal, K., Singhal, V., Brayton, R.: Verifying continuous time
  markov chains. In: International Conference on Computer Aided Verification.
  pp. 269--276. Springer (1996)

\bibitem{biere1999symbolic}
Biere, A., Cimatti, A., Clarke, E., Zhu, Y.: Symbolic model checking without
  bdds. In: International conference on tools and algorithms for the
  construction and analysis of systems. pp. 193--207. Springer (1999)

\bibitem{braitling2011counterexample}
Braitling, B., Wimmer, R., Becker, B., Jansen, N., {\'A}brah{\'a}m, E.:
  Counterexample generation for markov chains using smt-based bounded model
  checking. In: Formal Techniques for Distributed Systems, pp. 75--89. Springer
  (2011)

\bibitem{crn-analysis:cav2019}
{\v{C}}e{\v{s}}ka, M., K{\v{r}}et{\'i}nsk{\'y}, J.: Semi-quantitative
  abstraction and analysis of chemical reaction networks. In: Dillig, I.,
  Tasiran, S. (eds.) Computer Aided Verification. pp. 475--496. Springer
  International Publishing, Cham (2019)

\bibitem{chellaboina2009modeling}
Chellaboina, V., Bhat, S.P., Haddad, W.M., Bernstein, D.S.: Modeling and
  analysis of mass-action kinetics. IEEE Control Systems Magazine
  \textbf{29}(4),  60--78 (2009)

\bibitem{daigle2011automated}
Daigle~Jr, B.J., Roh, M.K., Gillespie, D.T., Petzold, L.R.: Automated
  estimation of rare event probabilities in biochemical systems. The Journal of
  chemical physics  \textbf{134}(4),  01B628 (2011)

\bibitem{dehnert2017storm}
Dehnert, C., Junges, S., Katoen, J.P., Volk, M.: A storm is coming: A modern
  probabilistic model checker. In: International Conference on Computer Aided
  Verification. pp. 592--600. Springer (2017)

\bibitem{gillespie2019guided}
Gillespie, C.S., Golightly, A.: Guided proposals for efficient weighted
  stochastic simulation. The Journal of chemical physics  \textbf{150}(22),
  224103 (2019)

\bibitem{han2007counterexamples}
Han, T., Katoen, J.P.: Counterexamples in probabilistic model checking. In:
  International Conference on Tools and Algorithms for the Construction and
  Analysis of Systems. pp. 72--86. Springer (2007)

\bibitem{han2007providing}
Han, T., Katoen, J.P.: Providing evidence of likely being on time:
  Counterexample generation for ctmc model checking. In: International
  Symposium on Automated Technology for Verification and Analysis. pp.
  331--346. Springer (2007)

\bibitem{hansson1994logic}
Hansson, H., Jonsson, B.: A logic for reasoning about time and reliability.
  Formal aspects of computing  \textbf{6}(5),  512--535 (1994)

\bibitem{kuwahara2008efficient}
Kuwahara, H., Mura, I.: An efficient and exact stochastic simulation method to
  analyze rare events in biochemical systems. The Journal of chemical physics
  \textbf{129}(16),  10B619 (2008)

\bibitem{kwiatkowska2007stochastic}
Kwiatkowska, M., Norman, G., Parker, D.: Stochastic model checking. In:
  International School on Formal Methods for the Design of Computer,
  Communication and Software Systems. pp. 220--270. Springer (2007)

\bibitem{kwiatkowska2011prism}
Kwiatkowska, M., Norman, G., Parker, D.: Prism 4.0: Verification of
  probabilistic real-time systems. In: International conference on computer
  aided verification. pp. 585--591. Springer (2011)

\bibitem{moura2008z3}
Moura, L.d., Bj{\o}rner, N.: Z3: An efficient smt solver. In: International
  conference on Tools and Algorithms for the Construction and Analysis of
  Systems. pp. 337--340. Springer (2008)

\bibitem{roberts2022stamina}
Roberts, R., Neupane, T., Buecherl, L., Myers, C.J., Zhang, Z.: Stamina 2.0:
  Improving scalability of infinite-state stochastic model checking. In:
  International Conference on Verification, Model Checking, and Abstract
  Interpretation. pp. 319--331. Springer (2022)

\bibitem{wimmer2009counterexample}
Wimmer, R., Braitling, B., Becker, B.: Counterexample generation for
  discrete-time markov chains using bounded model checking. In: International
  Workshop on Verification, Model Checking, and Abstract Interpretation. pp.
  366--380. Springer (2009)

\end{thebibliography}
\end{document}